\documentclass{amsart}
\usepackage{amsmath,amssymb,amsthm}
\usepackage{graphicx}
\usepackage{booktabs}
\usepackage[normalem]{ulem}
\newtheorem{theorem}{Theorem}
\newtheorem{cor}{Corollary}

\newcommand{\define}[1]{\emph{#1}}
\newcommand{\qsat}{\textsc{qsat}}
\newcommand{\sat}{\textsc{sat}}
\newcommand{\pspace}{\textsc{pspace}}
\newcommand{\prob}[1]{\textsc{#1}}
\DeclareMathOperator{\poly}{poly}

\begin{document}

\title{Multinational War is Hard}
\date{February 27, 2015}
\author{Jonathan Weed}
\address{MIT Department of Mathematics\\
50 Ames Street\\
Cambridge, MA 02139\\
USA}
\email{jweed@mit.edu}
\begin{abstract}
In this paper, we show that the problem of determining whether one player can force a win in a multiplayer version of the children's card game War is \pspace-hard. The same reduction shows that a related problem, asking whether a player can survive $k$ rounds, is \pspace-complete when $k$ is polynomial in the size of the input.
\end{abstract}
\maketitle

\section{Introduction}
War is a card game with extremely simple rules. At the beginning of each round, each player reveals the top card of her deck. The player with the best card wins all the revealed cards and places them at the bottom of her deck. Players are eliminated when they lose all their cards; the winner is the last player standing. The game's many detractors assert that players have no control over game's outcome, so it isn't worth playing. Nevertheless, the players do have one way in which to exert control: they get to choose the order in which captured cards are returned to the bottom of their deck. This small choice is enough to make the game provably difficult, though alas no more enjoyable to play.

Recent work on War has analyzed the question of how long a game of War can last. A stochastic model of a War-like game analyzed in~\cite{bennaim02} was conjectured to end in $O(n^2)$ rounds, but in fact, some games can last forever. Two recent papers independently exhibited an arrangement of a standard $52$-card deck that cycles when players use a fixed rule to return cards to the bottom of their decks~\cite{spivey10, laks12}. Since the length of a game of War is potentially unbounded, the problem of deciding whether a position is a win for Player 1 is likely not in \pspace. 

We show in this paper that the problem of deciding whether a position is a win for Player 1 is \pspace-hard. Deciding whether Player 1 can survive $k$ rounds is also \pspace-hard in general, and \pspace-complete when $k$ is polynomial in the size of the input---this bounded decision problem is the best completeness result we can hope for, for the reasons given above.

\section{Preliminaries}
We begin by briefly reviewing the relevant rules of War and the properties of the \pspace-complete problem \qsat, quantified Boolean satisfiability.

\subsection{Rules of War}
War is traditionally played by two players with a standard 52-card deck. In this paper, we consider a natural generalization to $m$ players with an $n$-card deck. Each card is labeled with a nonnegative number called its \define{rank}. A card of rank~$i$ beats a card of rank~$j$ if $i < j$. So, for instance, a card of rank~$2$ beats a card of rank~$3$, and no cards beat a card of rank $0$.

Play proceeds in rounds. At the beginning of each round, each player reveals the top card in her deck. If one player's card beats all the other players', then she wins all the revealed cards. If, on the other hand, two or more players reveal cards that are at least as good as all other cards on the table, then those players have a \define{battle}. Each player in a battle puts three more cards on the table and then reveals the fourth card. If one player's fourth card beats all the other players' fourth cards, she wins all the cards played during the round, including those from the battle. Otherwise, the players who revealed the best cards have another battle, and so on. This ``repeated battle'' dynamic will be used extensively in our reduction. 

Finally, we note that we imagine War in this context as a game of perfect information: every player knows the arrangement of every other player's deck at all times.

We will represent by \prob{MultiWar} the problem of deciding, given a initial ordering of cards in a game of multiplayer War, whether Player 1 has a winning strategy. The main result of this paper is the following.

\begin{theorem}
\prob{MultiWar} is \pspace-hard.
\end{theorem}

Moreover, if we denote by \prob{MultiWarSurive}-$k$ the problem of deciding whether Player 1 can survive $k$ rounds, we have the following corollary.

\begin{cor}
\prob{MultiWarSurvive}-$k$ is \pspace-complete for $k = n^{O(1)}$.
\end{cor}

The proof of these results will follow from the reduction given in Section~\ref{reduction}.

\subsection{Quantified Boolean satisfiability}
Quantified Boolean satisfiability (\qsat) is an analogue of \sat\ for the complexity class \pspace\ of problems decidable in polynomial space. A \qsat\ instance is a Boolean expression preceded by alternating quantifiers, such that every odd-numbered variable appears with an existential quantifier and every even-numbered variable appears with a universal quantifier. This problem also has a well-known interpretation as a two-player game, in which universal quantifiers are seen as choices made by an adversary. More information about \qsat\ and this ``game'' interpretation can be found in~\cite{ppad}.

In this paper, we will use a version of \qsat\ restricted to \textsc{cnf} formulas. This special case is also \pspace-complete and offers a natural starting point for our reduction.

\section{Reduction from \qsat} \label{reduction}
Given a \qsat\ instance with $m$ clauses and $2n$ variables, we will construct a War instance with $m + 6$ players and $\Theta(nm^2)$ cards. Our reduction will show that Player 1 has a satisfying strategy in \qsat\ if and only if she has a winning strategy in War. We first give a high-level overview of the reduction before giving a detailed example in Section~\ref{sec:ex}.

\subsection{Players and basic structure}
We will begin by explaining the roles of the players and the phases of the game. This information is summarized in Tables~\ref{players} and~\ref{phases}.
\newpage
\subsubsection{Players}
\begin{table}
\begin{center}
	\begin{tabular}{@{}lll@{}}\toprule
	Name & Abbreviation & Role \\ \midrule
	Players 1 \& 2 & P1 \& P2 & Choose variable assignments \\
	Clauses $1 \dots m$ & $\mathrm L1\dots \mathrm Lm$ & Give tokens to P1 when clause is satisfied \\
	Checker & Ch & Verify that P1 holds tokens for all clauses \\
	Cleanup & Cl & Infrastructure \\
	Parity [2] & * &  Infrastructure \\
	\bottomrule
	\end{tabular}
	\caption{Players}\label{players}
\end{center}
\end{table}
\begin{description}
\item[Player 1 (P1)] Player 1 is the player whose strategy we seek to determine. In our War instance, we will ensure that P1 only really has free choice at the very beginning of the game, when her moves correspond to setting \qsat\ variables. All other choices will be fixed if she wants to win the game. 

\item[Player 2 (P2)] Player 2 alternates with P1 in selecting values for variables at the beginning of the game. P2 will effect this choice by blocking P1 from setting a variable to a certain value, thereby forcing it to be assigned as P2 wishes. P2 will run out of cards and lose the game at the end of the Collection phase.
\item[Clause (L)] Each Clause player has an initial deck that encodes the literals of one clause in the instance of \qsat. If a clause contains a positive or negative literal corresponding to variable $x_i$, the corresponding Clause player will have a special card at the $(5n + (i-1)(m+6) + 1)$th or $(5n + (i-1)(m+6) + 2)$th position, respectively. Such a card is called a \define{token.} We will label tokens by capital letters $A, B, \dots$, and will assume for the purposes of ranking that $3 < A < B < \dots < 4$. Each clause contains copies of only one token type, and each token type is unique to a clause. If P1 is able to satisfy the \qsat\ instance via some assignment of variables, she will be able to collect at least one token of each type. 

L players lose at the end of the Collection phase, and they win no rounds.

\item[Checker (Ch)] The Checker player has no role until the Verification phase, until which point his deck contains only ``dummy'' cards which do not affect the outcome of any round. Ch contains copies of tokens of each type, and the only way for P1 to survive the Verification phase will be for her to do battle with Ch once for each token type. In this way, P1 will prove that she has collected copies of each token and has thereby satisfied the original \qsat\ instance. Ch will win many rounds during the Verification step, but his deck is long enough that the captured cards will not be played again until the Destruction phase, when Ch will be eliminated.

\item[Cleanup (Cl)] The Cleanup player's deck is seeded with powerful cards so that he can win rounds during the Collection, Sanitization, and Verification steps that we do not want affecting play later in the game. Cl also serves as a gatekeeper, whose powerful cards will eliminate P1 unless she does battle repeatedly with Ch during the Verification step. Finally, the end of Cl's deck will contain many powerful cards that he will lose to P1 over the course of the Destruction phase via a series of battles, thereby giving P1 enough cards to eliminate Cl and Ch and win the game.

\item[Parity (*)] There are two Parity players. They never win a round, and they exist only so there are five players left in the game during the Verification phase, after which they both run out of cards. 
\end{description}

\subsubsection{Phases} \label{sec:phases}
\begin{table}
\begin{center}
	\begin{tabular}{@{}ll@{}}\toprule
	Phase & Purpose \\ \midrule
	\oldstylenums{1}: Assignment & P1 and P2 choose values for each variable \\
	\oldstylenums{2}: Collection & P1 wins tokens from $\mathrm L$ players whenever clause is satisfied \\
	\oldstylenums{3}: Sanitization & P1 discards excess cards \\
	\oldstylenums{4}: Verification & Cl and Ch jointly verify that P1 has satisfied all clauses \\
	\oldstylenums{5}: Destruction & P1 wins enough powerful cards to win all remaining rounds \\
	\bottomrule
	\end{tabular}
	\caption{Phases}\label{phases}
\end{center}
\end{table}
Our instance of War proceeds in phases. The length of each phase in rounds is fixed and depends on the initial \qsat\ instance but does not change over the course of the game.

\begin{description}
\item[Assignment] In the first half of the Assignment phase, P1 and P2 win rounds in sets of four in the pattern P1, P2, P2, P1. Whenever P1 wins a round, she sets the value of a variable. At the first step of each set, that choice is unconstrained; at the fourth, that choice is fixed by the choice made by P2 at the third step. There are $n$ such sets, during which P1 sets all $2n$ variables.

The second half of the Assignment phase lasts $n$ rounds, and P1 wins all of them. These extra cards are ``buffers'' for the end of the Collection phase, and their purpose will be discussed in more detail below.

The Assignment phase ends when P1's deck wraps around for the first time.

\item[Collection] P1 wins exactly $2n$ rounds during the Collection phase, but which rounds she wins depends on the choices made during the Assignment phase. Each round P1 wins corresponds to a literal, and she wins tokens corresponding to the clauses containing that literal. There are also many rounds that do not correspond to literals, but Cl holds powerful cards in these positions, so P1 does not win any such clauses. The L players lose at the end of this step, once P1 has claimed all possible clause tokens. 

P2 also wins $n$ rounds during the Collection phase, but we do not want P2 to survive into the Sanitization phase. The Collection phase therefore ends with a string of rounds won by Cl, who holds a run of consecutive unbeatable cards of exactly the length of P2's deck. P2 loses all these rounds and is thereby eliminated, but P1 holds buffer cards at these locations, so her important cards are not touched.

The Collection phase ends when P1's deck wraps around for the second time, and P2 is eliminated. At this point, five players remain.

\item[Sanitization] P1's deck now contains $2n(m+6)$ cards, of which only some are valuable. The Sanitization phase will give her the chance to discard unwanted cards. Five players remain (P1, Ch, Cl, *, *), and during the first half of the Sanitization phase all players except P1 hold very weak cards. P1 wins one round for each card in her deck. When she places the set of five cards won in a round at the bottom of her deck, she can either keep her original card on top or move it elsewhere in the set of five cards.

In the second half of the Sanitization phase, P1 passes through her deck one more time. Cl's deck contains sets of five cards, where the first card in each set is moderately powerful and the last four cards are unbeatable. Token cards will beat the first card of each set, so P1 can preserve a desired token by placing it first in a set of five cards. If she places it elsewhere, it will be eliminated.

In order to survive the Verification round, P1 will need to hold $5m+5$ cards: $m$ unique tokens, $1$ sentinel card of rank $3$, and $4m + 4$ dummy cards spaced evenly between them. If P1 has managed to collect tokens of all types during the Collection phase, she will be able to use the Sanitization phase to set her deck appropriately before Verification. 

The Verification phase ends when P1's deck wraps around for the fourth time. 

\item[Verification] In the Verification phase, Cl and Ch jointly check that P1 has one token of each type. The Verification phase comprises $m$ sub-phases, each one designed to check for the presence of a specific token. The tokens are checked from least to most powerful; at the end of each sub-phases, the token being checked is removed from P1's deck and the rest of her cards are left essentially undisturbed.

The sub-phases are separated by long runs of very powerful cards in the decks of both Ch and Cl, called \define{the gauntlet}. These runs come in alternating sets of four, and initially Ch and Cl's sets are out of sync with each other so that none of their cards match. The threat of defeat is omnipresent during the Verification round, since each sub-phase could mean the end of P1 if she does not somehow bypass the gauntlet each time. She will do so by causing a battle with Ch. A battle forces P1 and Ch alone to play four additional cards. As a result, Cl and Ch's decks shift relative to each other by four cards, and the alternating sets of the gauntlet will align. When the gauntlet is reached, therefore, Ch and Cl will do battle many times before Ch eventually wins all the cards in the gauntlet without P1 having ever had to beat them. Then next sub-phase then begins. 

At the end of the Verification phase, the Parity players run out of cards and lose. This phase ends when all $m$ tokens have been checked, and P1's deck wraps around for the $(m+4)$th time.

\item[Destruction] If P1 has managed to survive the Verification phase, then her deck contains exactly five cards: the sentinel card, three dummy cards, and one copy of the $A$ token. If she arranges the cards in that order (which she can do when she wins the last round of the Verification phase), she will capture enough powerful cards during the Destruction phase to win the game. During the Destruction phase, Cl has a series of battles with P1, during which P1 wins a great number of unbeatable cards. A battle allows a player with weaker cards to win stronger cards if those stronger cards appear as part of the set of three cards each player ``wagers'' before turing up a fourth card. Since the contents of P1's deck are entirely knowable, we can arrange for Cl's cards to match P1's repeatedly as P1 wins more and more battles. Each time she does so, her deck gets larger and eventually contains a run of cards long enough to eliminate Cl and Ch entirely. If P1 enters this round, therefore, she can successfully win the game.

\end{description}

\subsubsection{Bounds on size and length of War instance}

The general outline above is detailed enough to make some comments about the size and length of our instance of War. We recall that the \qsat\ instance contains $2n$ variables and $m$ clauses, and note that the length of the game does not depend on choices made by P1, except when those choices would cause P1 to lose prematurely.

The Assignment phase lasts $5n$ rounds, and this is also the initial length of P1's deck. At the end of this phase, P1 holds $3n(m+6)$ cards.

The Collection phase lasts $3n(m+6)$ rounds, one for each card in P1's deck. Of these, only $4n$ rounds are meaningful, insofar as they represent genuine choices P1 can make. The other rounds are won by either Cl or P2 and do not contribute to rest of play. P1 wins exactly $2n$ rounds during this phase, and ends the phase holding $2n(m+6)$ cards.

The Sanitization phase lasts $12n(m+6)$ rounds, where P1 wins all of the first $2n(m+6)$ rounds and loses all but $m+1$ of the last $10n(m+6)$ rounds. At the end of the round, P1 holds $5m+5$ cards.

The $i$th sub-phase of the Verification phase lasts $5(m+2-i)$ rounds, with the final round of each sub-phase involving a long battle between Ch and Cl involving at least $8m$ cards from each player. The total number of rounds of this phase is $5(m^2 + 3m)/2$, and it consumes at least $11m^2$ cards from Ch and Cl's decks. At the end of this phase, P1 holds $5$ cards, and Ch and Cl have won at most $4nm^2$ cards during the course of the game.

In the first half of the Destruction phase, P1 must win enough cards to be able to win all remaining rounds of the game. This requires that P1 hold at least $4nm^2$ powerful cards. Battles allow P1 to effectively double the size of her deck every round during this half of the destruction phase, so winning the necessary cards takes $O(\log n + \log m)$ rounds. Finally, the second half of the Destruction phase lasts at most $4nm^2$  rounds before Ch and Cl are eliminated and the game ends. 

The total number of required rounds is therefore $O(nm^2)$, and this is also a bound for the total number of required cards. In particular, the reduction requires space and time  polynomial in the size of the input. 

If P1 survives the Verification phase, she has demonstrated that she can satisfy the \qsat\ instance, so deciding whether P1 survives to the beginning of the Destruction phase is also \pspace-hard. Moreover, the number of rounds before the beginning of the Destruction phase is a number $k = \poly(m, n)$ depending only on the size of the input. Then the problem \prob{MultiWarSurvive}-$k$ is in \pspace\ and therefore is \pspace-complete. 

\subsection{A detailed example} \label{sec:ex}
Because of the large number of cards required to represent even a simple instance of \qsat, it is impractical to give a full example of this reduction except in trivial cases. In this section, we will do just that: though our \qsat\ instance is small, the War instance we produce will contain all gadgets necessary to produce an instance in the general case. 

The \qsat\ instance we consider will be given by the following formula:
\begin{equation*}
\exists x_1 \forall x_2 (x_1) \wedge (x_1 \vee \neg x_2).
\end{equation*}
We note that this \qsat\ instance is satisfiable, and so our reduction produces a winning game for P1.

\subsubsection{Assignment}

\begin{table}
\begin{center}
	\begin{tabular}{@{}lll@{}}\toprule
	P1 & P2 & Notes\\ \midrule
	2 & x & P1 choice gadget \\
	x & x & \\
	\cmidrule(r{1em}){1-2}
	x & 1 & P2 choice gadget \\
	2 & x & \\
	\cmidrule(r{1em}){1-2}
	2 & x & Buffer \\
	\bottomrule
	\end{tabular}
	\caption{Starting decks}\label{choice}
\end{center}
\end{table}

Table~\ref{choice} shows the starting decks for P1 and P2. We adopt the very useful convention that the letter x represents an unspecified weak card, with the promise that no x card will ever win a round. This allows us to better focus on the role of important cards in the reduction. Whenever players are omitted from a listing of decks, as in Table~\ref{choice}, we assume that their decks contain x's unless otherwise specified.

P1 wins the first round in the P1 choice gadget, thereby winning a set of cards $\{2, \mathrm x, \mathrm x, \dots, \mathrm x\}$. If P1 places the 2 card at the top of this set, then we say she sets the corresponding variable to true. If she places the 2 card at the second position, she sets it to false. Any other position for the 2 card does not correspond to a literal and has no bearing on the rest of the game. (In particular, placing the 2 card elsewhere does not help her.) P2 then wins a round, so that both players have won the same number of cards and therefore stay in sync.

P2 wins the first round in the P2 choice gadget and wins the set $\{1, \mathrm x, \mathrm x, \dots, \mathrm x\}$. P2 also has the choice of putting the 1 card first or second; in this case, though, the position he chooses for the 1 card should be \emph{opposite} the value he seeks to set: putting it first corresponds to ``blocking'' true (thereby setting the variable to false), and putting it second blocks false (thereby setting the variable to true). When P1 wins the second hand in the choice gadget, she wins the set $\{2, \mathrm x, \mathrm x, \dots, \mathrm x\}$. If she lines the 2 card up with the 1 card, she wins no rounds corresponding to literals; if she does not, she can win rounds corresponding to the variable value that P2 has not blocked.

Finally, P1 wins the only round in the Buffer gadget, and her choice of how to arrange these cards has no bearing on the game.

For concreteness, we will suppose that the players set both $x_1$ and $x_2$ to true.

\subsubsection{Collection}

\begin{table}
\begin{center}
	\begin{tabular}{@{}llllll@{}}\toprule
	P1 & P2 & L1 & L2 & Cl & Notes\\ \midrule
	2 & x & A & B & 3 & $x_1$ \\
	x & x & x & x & 3 & $\neg x_1$ \\
	x & x & x & x & 1 &  \\
	x & x & x & x & 1 &  \\
	x & x & x & x & 1 &  \\
	x & x & x & x & 1 &  \\
	x & x & x & x & 1 &  \\
	x & x & x & x & 1 &  \\
	\cmidrule(r{1em}){1-5}
	2 & x & x & x & 3 & $x_2$ \\
	\textbf{x} & \textbf 1 & \textbf x & \textbf B & \textbf 3 & $\neg x_2$ \\
	x & x & x & x & 1 &  \\
	x & x & x & x & 1 &  \\
	x & x & x & x & 1 &  \\
	x & x & x & x & 1 &  \\
	x & x & x & x & 1 &  \\
	x & x & x & x & 1 &  \\
	\cmidrule(r{1em}){1-5}
	2 &  & x & x & 0 &  Cleanup \\
	x &  & x & x & 0 &  \\
	x &  & x & x & 0 &  \\
	x &  & x & x & 0 &  \\
	x &  & x & x & 0 &  \\
	x &  & x & x & 0 &  \\
	x &  & x & x & 0 &  \\
	x &  & x & x & 0 &  \\
	\bottomrule
	\end{tabular}
	\caption{Collection phase}\label{collection}
\end{center}
\end{table}

The Clause gadgets appear in Table~\ref{collection}, which depicts the entirety of the Collection phase. Note that most rounds in this phase do not carry any particular meaning and are won automatically by Cl. P2's deck at the beginning of this phase is shorter than P1's deck, but P2 wins the bolded round and these cards (along with P1's buffer cards) are captured by the Cleanup gadget. P2, L1, and L2 run out of cards at the end of this phase.

P1 wins tokens from both L1 and L2, corresponding to the fact that setting $x_1$ to true satisfies both clauses in the initial instance of \qsat. At the end of this phase, her deck contains A and B tokens as well as other cards that will not be necessary for the remainder of the reduction. In order to prepare herself for the Sanitization phase, P1 should make sure the last card in her deck is a $2$ card; the arrangement of all other cards is arbitrary.

\subsubsection{Sanitization}
The gadgets for the Sanitization phase appear in Table~\ref{sanitization}, where we use a question mark symbol to indicate that the order of P1's cards does not matter. The cards marked y in Cl's deck indicate that we want P1 to win each of the first $2n(m+6)$ rounds, so during this period no player should hold cards more powerful than anything P1 has. We required that P1's last card at the end of the Collection phase be a 2 card so that she can win a 3 card to use as a sentinel.

P1's goal should be to use the second half of the Sanitization phase to put her deck in the correct form for the Verification phase; in particular, she should aim to preserve exactly one copy of each token as well as one sentinel card of rank 3. She can accomplish this by shifting cards she wishes to preserve into the ``windows'' that appear in Cl's deck every five rounds during the second half of the phase. Any token that she places in a window is preserved in the next phase. 

\begin{table}
\begin{center}
	\begin{tabular}{@{}lll@{}}\toprule
	P1 & Cl & Notes\\ \midrule
	? & y & $2n(m+6)-1$ rounds \\
	2 & 3 & $1$ round \\
	\cmidrule(r{1em}){1-2}
	? & 4 & $10n(m+6)$ rounds \\
	? & 1 & \\
	? & 1 & \\
	? & 1 & \\
	? & 1 & \\
	\bottomrule
	\end{tabular}
	\caption{Sanitization phase}\label{sanitization}
\end{center}
\end{table}

\subsubsection{Verification}
One sub-phase of the Verification phase is presented in Table~\ref{verification}. Note that P1's deck is of the desired form. The first sub-phase checks for the presence of a B token, and the second checks for the presence of an A token. (In general, there will be $m$ sub-phases, one for each token.) The function of the gauntlet is described in Section~\ref{sec:phases}. 

\begin{table}
\begin{center}
	\begin{tabular}{@{}llll@{}}\toprule
	P1 & Ch & Cl & Notes\\ \midrule
	B & B & x & B checker\\
	x & 1 & x & \\
	x & 1 & x & \\
	x & 1 & x & \\
	x & 1 & x & \\
	\cmidrule(r{1em}){1-3}
	A & B & x & B checker\\
	x & 1 & x & \\
	x & 1 & x & \\
	x & 1 & x & \\
	x & 1 & x & \\
\cmidrule(r{1em}){1-3}	
	3 & C & 2 & Gauntlet \\
	x & 1 & 2 & \\
	x & 1 & 2 & \\
	x & 1 & 2 & \\
	x & 2 & 1 & \\
	  & 2 & 1 & \\
	  & 2 & 1 & \\
	  & 2 & 1 & \\
	  & \multicolumn{2}{c}{\dots} \\	  	  
	\bottomrule
	\end{tabular}
	\caption{Verification sub-phase}\label{verification}
\end{center}
\end{table}

\subsubsection{Destruction}

At the end of the Verification phase, P1 holds five cards: a 3 card, three x cards, and an A card. Table~\ref{destruction} shows the first two stage of Destruction, in which Cl  gives P1 a large number of 0 cards---the most powerful cards in the game. We require at the end of the first stage that P1 place  captured 0 cards at the first and fifth positions in her stack, so that she continues to cause battles with Cl in the second stage. Continuing in this way, Cl can feed enough 0 cards to P1 so she can win the game.

\begin{table}
\begin{center}
	\begin{tabular}{@{}lll@{}}\toprule
	P1  & Cl & Notes\\ \midrule
	3 & 3 & Stage 1\\
	x & 0 & \\
	x & 0 & \\
	x & 0 & \\
	A & x & \\
	\cmidrule(r{1em}){1-2}
	  & 0 & Stage 2 \\
	  & 0 & \\
	  & 0 & \\
	  & 0 & \\
	  & 0 & \\
	  & 0 & \\
	  & 0 & \\
	  & 0 & \\	
	  & x & \\  	  	  	  	  	  
	  & \dots & \\	  	  
	\bottomrule
	\end{tabular}
	\caption{Destruction phase}\label{destruction}
\end{center}
\end{table}

\section{Conclusion}

We have shown via a reduction from \qsat\ that \prob{MultiWar} is \pspace-hard. There are still many questions left unanswered by this work, however. In particular, the question of whether the game of two-player war is \pspace-hard is still open. Our reduction relied heavily on the fact that we had an unbounded number of players, and very few of our constructions seem easy to implement in a two-player setting. 

The finiteness of War is still not well understood, even in the two-player setting. The problem of deciding whether a certain player can force a configuration to cycle is apparently still open. Moreover, deciding whether a certain configuration can cycle given \emph{some} choice of strategy seems equally difficult except in certain trivial cases. 

In another direction, given that \prob{MultiWar} is probably not in \pspace, it may be possible to show that \prob{MultiWar} is hard with respect to some larger complexity class. These open problems suggest several avenues for future research and show that the story of War's computational complexity is far from over.

\bibliographystyle{amsplain}
\bibliography{final_project}

\section*{Acknowledgements}
I would like to thank Erik Demaine, Jayson Lynch, and Sarah Eisenstat for their inspiration and assistance. In particular, I would like to thank Sarah for suggesting that I focus on multiplayer War.

\end{document}